\documentclass[10pt]{article}
\usepackage{graphicx}
\usepackage{amssymb,amsmath,dsfont}
\usepackage{float}
\usepackage{hyperref}
\hypersetup{
	colorlinks=true,
	linkcolor=blue,
	filecolor=magenta,      
	urlcolor=cyan}

\begin{document}

\title{Efficient algorithms for the dense packing of congruent circles inside a square}
\author{Paolo Amore \\
{\small Facultad de Ciencias, CUICBAS, Universidad de Colima},\\
{\small Bernal D\'{i}az del Castillo 340, Colima, Colima, Mexico} \\
{\small paolo@ucol.mx} \\
Tenoch Morales  \\
	\small Facultad de Ciencias, Universidad de Colima,\\
	\small Bernal D\'{i}az del Castillo 340, Colima, Colima, Mexico \\
	\small tenochmorales0@gmail.com }

\maketitle

\begin{abstract}
We study dense packings of a large number of congruent non-overlapping circles inside a square by looking for configurations  which maximize the packing density, defined as the ratio between the area occupied by the disks and the area of the square container. 
The search for these configurations is carried out with the help of two algorithms that we have devised: a first algorithm is in charge of obtaining sufficiently dense configurations  starting from a random guess, while a second algorithm improves the configurations obtained in the first stage.
The algorithms can be used sequentially or independently.

The performance of these algorithms is assessed by carrying out numerical tests for configurations with a large number of circles.
\end{abstract}

\section{Introduction}
\label{sec:intro}

In this paper we consider the problem of packing $N$ congruent circles inside a square of side $L$. We ask what is the maximal density $\rho$ that can be achieved if the circles are not allowed to overlap but they can be in contact with each other or with the border of the square. 

The density of the packing can be expressed as
\begin{equation}
\rho =  \frac{A_{\rm circles}}{A_{\rm square}} =  \frac{N \pi r_0^2}{L^2} \ ,
\end{equation}
where $r_0$ is the radius of the circles and it is bound by the maximal density for circle packing in the plane~\cite{FejesToth42}
\begin{equation}
\rho \leq \rho_{\rm plane} \equiv \frac{\pi}{\sqrt{12}} \approx 0.9069 \ .
\end{equation}

Thus the quest for optimality amounts to finding either the largest $r_0$ for a square of given $L$, or the smallest $L$ for $N$ given circles of radius $r_0$.  Refs.~\cite{Schaer65,Schaer65b} by Schaer and Schaer \& Meir are the first papers where optimal solutions for packing of equal circles inside a square are discussed ($9$ and $8$ circles respectively), although Schaer mentions that the optimal solution for $6$ circles had been found earlier by R.L. Graham (unpublished). Over the years, several researchers have studied this problem, finding the optimal solutions up to $N = 33$~\footnote{The optimal solutions for $N=31,32,33$ have been proved recently by Mark\'ot in \cite{Markot21}. } (a partial list of references can be found in Table 1.3 of ~\cite{Szabo07b}). The case $N=36$, which corresponds to a square packing, also appear to be optimal.
In parallel, there has also been an effort to obtain approximate packings, without a proof of optimality, using numerical methods (see also  Table 1.3 of ~\cite{Szabo07b}).  It is also worth mentioning \cite{Hifi09} contains a recent literature review 
on circle and sphere packing problems. In addition to these references, E. Specht has created a digital repository that contains approximate configurations for packing of circles (not necessarily being identical) in different domains  \cite{SpechtRepo}. This page contains a large number of configurations, obtained by different researchers and it constitutes a remarkable source of information for anyone interested in circle packing: for the square, the largest configuration reported in  \cite{SpechtRepo} contains $9996$ circles (it must be said, however, that for very large $N$ many of the configurations reported in \cite{SpechtRepo} are most likely not optimal~\footnote{Notice for example that the density reported for $N=2000$ is smaller than the density reported for $N=500$.}). Not surprisingly, finding high density configurations rapidly becomes a very demanding task for $N \gg 1$.

Other domains for which circle packing has been studied are the  circle, equilateral and isosceles triangles and rectangles of different proportions (see ~\cite{Szabo07b,SpechtRepo} for the relevant bibliography). A full account of the previous literature on this subject  goes beyond our possibilities and the reader interested in this topic should refer to ~\cite{Szabo07b,Hifi09,SpechtRepo} and references therein. The purpose of our paper is to focus on the specific problem of circle packing inside a square and  to describe an alternative method to obtain dense configurations, by extending a method previously devised by Nurmela and  Östergård in ~\cite{Nurmela97}~\footnote{The ideas at the base of our algorithm, however, can be generalized to containers of different shape.}.

The general problem of packing $N$ circles inside a finite domain on the plane is interesting under different perspectives: to a general audience, it may appeal because it is an example of a problem that one encounters in everyday life, trying to adjust a number of (not necessarily rigid) objects in the smallest area (volume) possible (in this case, circle packing may constitute an excessive idealization of a realistic situation); to a different scale the problem is relevant in industrial and commercial applications, where fitting the largest number of equal objects in a minimal space brings an economical advantage (additionally, one may think of the situation in which a maximal number of objects needs to be produced out of a given quantity of material, something that must present often to a tailor or a carpenter, just to mention two examples); to mathematicians, the question of finding optimal arrangements (packings) is interesting per se, particularly taking into account the apparent simplicity of the problem and the difficulty in reaching rigorous results~\footnote{Kepler's conjecture, regarding sphere packing in three--dimensional Euclidean space, has been proved only in recent times by Thomas Hales~\cite{Hales05}, despite being put forward by Kepler in 1611. Needless to say, that the optimal configuration has been empirically known to generations of greengrocers without a formal training in mathematics!}; to physicists, this problem is relevant in describing the behavior of real physical systems, with short range (contact) interactions between the components. One of such applications is the study of the properties of granular materials~\cite{Stokely03,Salerno18,deBono20}. Additionally, it must be mentioned that circle packing is a NP hard optimization problem~\cite{Hifi09}.

The paper is organized as follows: in Section \ref{sec:method} we describe the algorithms; in Section \ref{sec:results} we apply the algorithms to calculate dense configurations for selected numbers of disks (particularly for large $N$ values we are able to improve the results reported in \cite{SpechtRepo}); finally in Section \ref{sec:conclusions} we resume our findings and briefly discuss the possible directions of future work.

\section{The method}
\label{sec:method}

In this section we will present our computational method for the calculation of dense packings of $N$  congruent circles in a square.  The method relies on two algorithms that can be used either sequentially or independently, depending on the needs.  We proceed now with a detailed discussion of these algorithms.

\subsection{Algorithm 1}
The first algorithm is an evolution of the packing algorithm first introduced by Nurmela and Östergård in \cite{Nurmela97}.

We will first describe the essence of their method and then illustrate ours. The goal is to determine the maximum radius of $N$ congruent disks that can be fitted inside a square of fixed side length; the disks are allowed to be in contact, both with other disks and with the border of the square, but must not overlap. Calling $2 r_{0}$ the minimal  distance between any two disks and $\ell$ the side of the  square in which the centers of the disk are confined (therefore the disks are contained inside a square of side $\ell+2 r_0$), the corresponding packing density will be
\begin{equation}
\rho = \frac{N \pi r_0^2}{(\ell + 2 r_0)^2} \ .
\end{equation}

\begin{figure}
	\begin{center}
		\includegraphics[width=4cm]{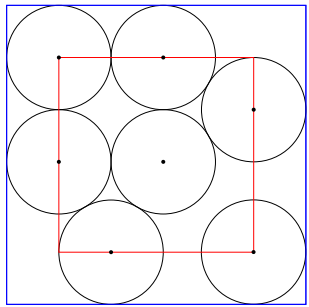} 
	\end{center}
	\caption{Optimal configuration of $7$ circles inside a square: the internal (red) square  has side length  $\ell = 1$, whereas the external (blue) square  has sidelength $L = 1+2r_0$. In this case $r_0 = 2 -\sqrt{3} \approx 0.267949$ and $\rho =  \frac{7}{169} \left(19-8 \sqrt{3}\right) \pi \approx 0.669311$}
	\label{Fig_0}
\end{figure}

The method of ~\cite{Nurmela97} is based on the minimization of the ``energy" function 
\begin{equation}
V = \sum_{i=2}^N \sum_{j=1}^{i-1} \left( \frac{\lambda}{r_{ij}^2} \right)^s \ ,
\label{eq_V}
\end{equation}
where $r_{ij}$ is the Euclidean distance between the centers of two circles and $\lambda$ is
a parameter introduced to avoid numerical overflows. The exponent $s>0$ is a real parameter that 
determines the short and long range behavior of the potential.  
For $s=1/2$, for instance, eq.~(\ref{eq_V}) is the total electrostatic energy of $N$ equal charges, confined in a given domain. The minimization of this function, particularly for the case where the  domain is the surface of a sphere, is known as {\sl Thomson's problem} and it has been studied by several authors \cite{Erber91,Perez96,Bowick02,Wales06,Wales14}.
However, the Coulomb interaction is long-range and therefore it is not suited for the problem of packing, where the circles interact only when they are in contact. Indeed, the appropriate energy function for packing should have $s \gg  1$.

The first step taken in \cite{Nurmela97} is to transform the original {\sl constrained} optimization 
problem, corresponding to finding a minimum of the potential $V$ of eq.~(\ref{eq_V}), to a {\sl unconstrained}
optimization problem, which  is easier to solve. This task is achieved  by expressing the Cartesian coordinates $(x_i, y_i)$ of the centers of the circles as
\begin{equation}
(x_i,y_i) = \frac{\ell}{2} (\sin t_i, \sin u_i)  \hspace{1cm} -\pi/2 \leq t_i, u_i \leq \pi/2 \ , 
\end{equation}
where $i =1, 2, \dots, N$ and $\ell$ is the side length of a square containing the centers of the circles (in the following 
we will work with $\ell=1$ without loss of generality). 

The transformed problem can be attacked by assigning the initial positions of the circles randomly (or according to a suitable criteria) and then proceeding with the minimization of the energy function corresponding to an appropriate value of $s$.  As we mentioned earlier, finding a good packing requires $s \gg 1$, but if $s$ is chosen too large at the beginning of the calculation, the interaction between points will be too weak and the resulting configuration will not be dense.  On the other hand,  if $s$ is too small, the interaction is long-range, with a single point being able to interact with {\sl all} remaining points; in this case, the center of the box will be depleted (a point in this region will have more neighbors and it will be convenient to move it toward the border) and the corresponding packing  density will be low. For this reason, Nurmela and Östergård start with a moderately small value of $s$, say $10 \leq s \leq 100$, finding the minimum of the corresponding energy function. We call $s_{\rm in}$ the initial value of $s$. Once that minimum is obtained the process is repeated, using  the configuration just obtained as initial guess and  making the change $s \rightarrow 2s$ inside the energy function. The process is iterated many times, until convergence is reached,  typically up to values of $s \approx 10^6$.  During these steps the parameter $\lambda$ is set equal to the square of the minimal distance between any two circles, and it is recalculated from time to time to avoid numerical instabilities caused by working with exceedingly small or large numbers. The packings obtained in this way are then further improved by identifying the disks that are possibly touching other disks or the border of the square; in this way one obtains a system of nonlinear equations that can be solved numerically.

We now describe our algorithm, starting with a key observation: an efficient optimization requires to take into account the different nature of the
points that do not touch the border of the square (internal points) and those that are in contact with the border of the square (border points). At equilibrium, the resultant of the forces acting on each of the internal points must vanish, whereas for the  border points only the component tangent to the border vanishes. For this reason, if a point has reached  the border at some stage of the calculation it is extremely unlikely that it can be moved inside afterwards. In fact, the configurations will tend to maintain or increase the number of border points as the iterations proceed~\footnote{A similar problem was observed by one of us, while studying the Thomson's problem in a disk~\cite{Amore16}: in that case, it was found that the energies of the configurations have a strong dependence on the number of border points and that the search for a global minimum of the total energy requires to generate configurations with the appropriate number of border points.}. 
In the original algorithm in \cite{Nurmela97} there is no mechanism preventing the internal points from  being deposited 
on the border if $s$ is small enough and this explains why the authors start their calculation with a rather large value of 
$s$.

To avoid these complications we modify the energy functional (\ref{eq_V}) to
\begin{equation}
V = \sum_{i=2}^N \sum_{j=1}^{i-1} \left( \frac{\lambda}{r_{ij}^2} \right)^s \mathcal{F}_{ij}(\epsilon,\alpha) \ , 
\label{eq_V2}
\end{equation}
with
\begin{equation}
\begin{split}
\mathcal{F}_{ij}(\epsilon,\alpha) &\equiv \left[\left( 1+\epsilon - \sin^2(t_i) \right) 
\left( 1+\epsilon - \sin^2(t_j) \right) \right.  \\
&\left. \cdot \left( 1+\epsilon - \sin^2(u_i) \right) \left( 1+\epsilon - \sin^2(u_j) \right) \right]^\alpha \ ,
\hspace{0.5cm} -\pi/2 \leq t_i, u_i \leq \pi/2
\end{split}
\label{eq_F}
\end{equation}
and $\epsilon>0$ and $\alpha \leq 0$. 

Observe that for $\alpha = 0$, the function (\ref{eq_V2}) reduces to (\ref{eq_V}), while for $\alpha<0$, 
it is energetically favorable to move the points away from the border of the square; in this way, configurations with a low border occupancy can be obtained even for modest values of $s$. 

Starting from an initial random configuration, for a sufficiently small value of $s$,  we iterate the basic algorithm of Nurmela and Östergård (NÖ), applied to the new  functional (\ref{eq_V2}), and progressively relax  the border repulsion as $s$ increases. Specifically we  
use $\alpha = -1/s$ and keep $\epsilon$ fixed at some small positive value (for example, $\epsilon \approx 10^{-10}$)~\footnote{Of course, different choices for $\alpha$ could be considered, keeping in mind that we want $\mathcal{F}$ to provide sufficient border repulsion at the start of the process and 
$\mathcal{F} \rightarrow 1$ at the end of the process.  }.
Given the higher computational cost of working with the functional  (\ref{eq_V2}), 
and taking into account that the effect of $\mathcal{F}$  is negligible when $s$ is large enough,
one can switch to the original  functional (\ref{eq_V}) in this case (typically we use $s > 10^3$).

The algorithm is composed of the following steps:
\begin{itemize}
	\item[1)] Generate a random set of $N$ points inside a square of side $\ell=1$;
	\item[2)] Define the energy functional (\ref{eq_V2}), with $s=s_{\rm in}$ and with $\alpha(s)$ such that $\lim_{s\rightarrow \infty} \alpha(s)=0$ (for instance $\alpha(s)=-1/s$); the initial value of $s$, $s_{\rm in}$, may be chosen arbitrarily, but it is convenient that it is not too large, to allow better results;
	\item[3)] Minimize the functional in step 2) and use the configuration obtained as new initial configuration, now with $s' = \kappa s$, with $\kappa>1$ (typically we use $\kappa \approx 1.5$, or $\kappa \approx 2$, as done in \cite{Nurmela97} ); 
	\item[4)] Repeat step 3) until reaching a large value of $s$ (in most of our calculations we have used $s_{\rm fin} = 10^6$, but larger values might be used if convergence was not reached);
	\item[5)] Compare the density of the configuration obtained at step 4) with the best density for $N$ circles obtained in earlier iterations and, if the current density improves the previous record, store the new configuration;
	\item[6)] Repeat steps 1-5) a large number of times (for $N \leq 100$, one can easily perform $1000$ trials in a limited time).
\end{itemize}

\begin{figure}
	\begin{center}
\includegraphics[width=4cm]{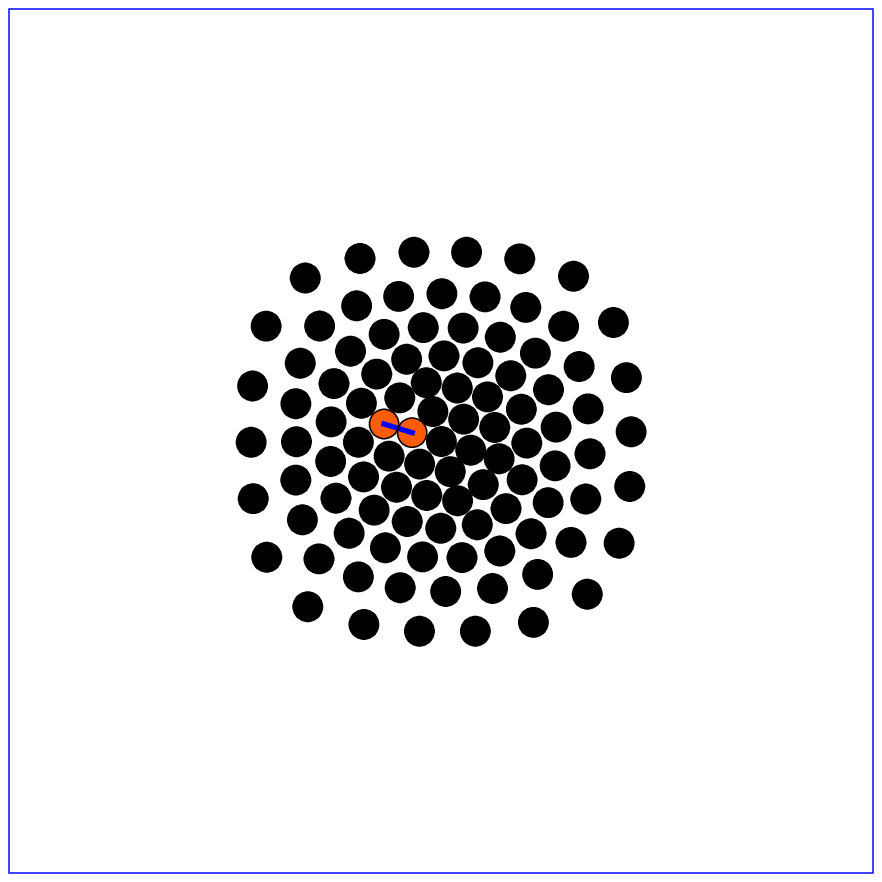} \hspace{1cm} \includegraphics[width=4cm]{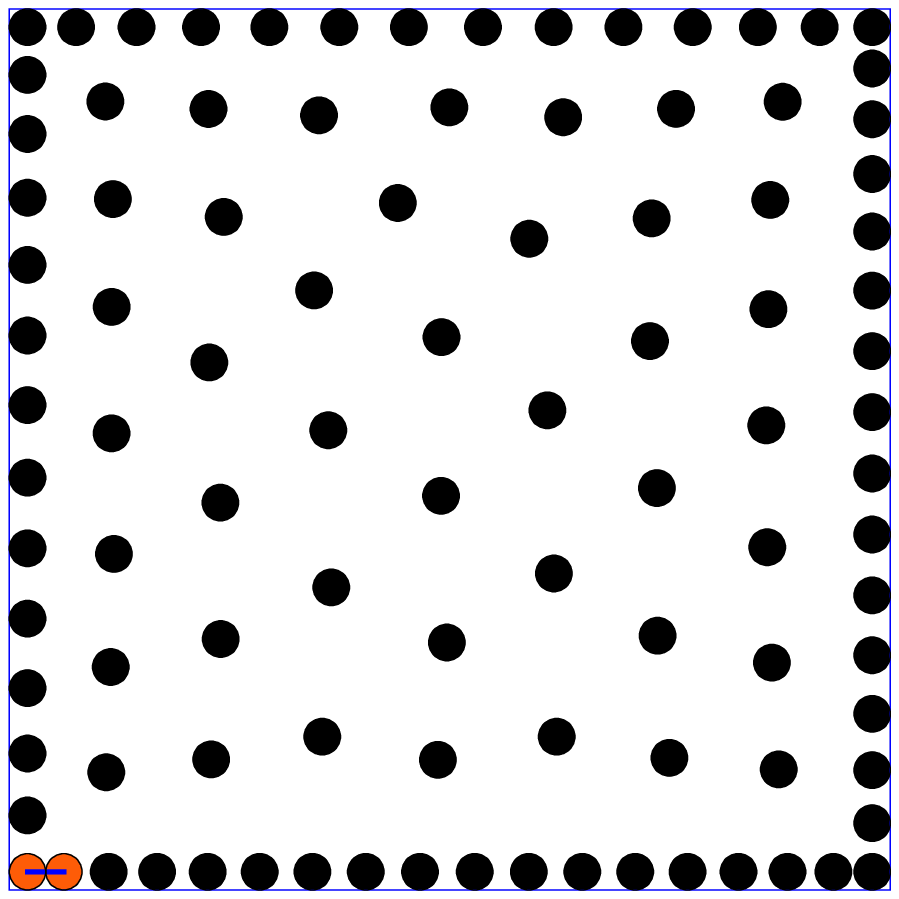}
	\end{center}
	\caption{Example of configurations of $N=100$ circles obtained for $s=1/2$ starting from the same initial random arrangement using the functional of eq.~(\ref{eq_V2}) (left figure) and of eq.~(\ref{eq_V}) (right figure). The disks at a minimal distance are colored.    }
	\label{Fig_1}
\end{figure}

We briefly discuss some of the features of the algorithm, with the help of figures and numerical tests.

In Fig.~\ref{Fig_1} we display two configurations of $N=100$ circles obtained for $s_{\rm in}=1/2$, starting from the same initial random arrangement, using either the functional of eq.~(\ref{eq_V2})  (left figure) or of eq.~(\ref{eq_V})  (right figure). 
In this second case  the majority of the circles  is on the border of the square: if this configuration is evolved to larger and larger values of $s$, the maximal packing density will then be determined by the side of the square with the largest number of circles.
On the other hand, the modification introduced with the functional (\ref{eq_V2}) allows one to start the algorithm with lower  values of $s_{\rm in}$, for which the interaction between points is larger, without increasing the number of border points.

\begin{figure}
\begin{center}
\includegraphics[width=5cm]{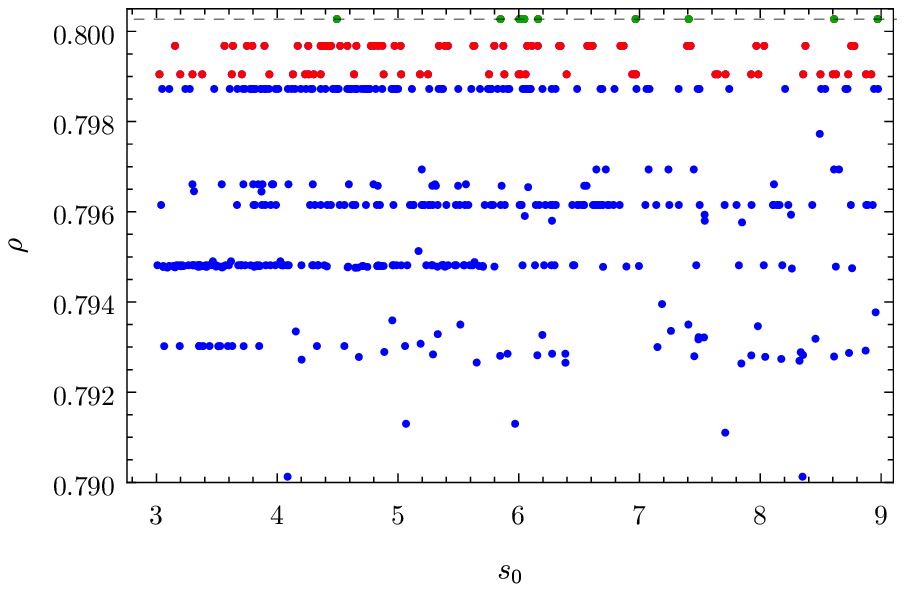} \hspace{0.5cm}
\includegraphics[width=5cm]{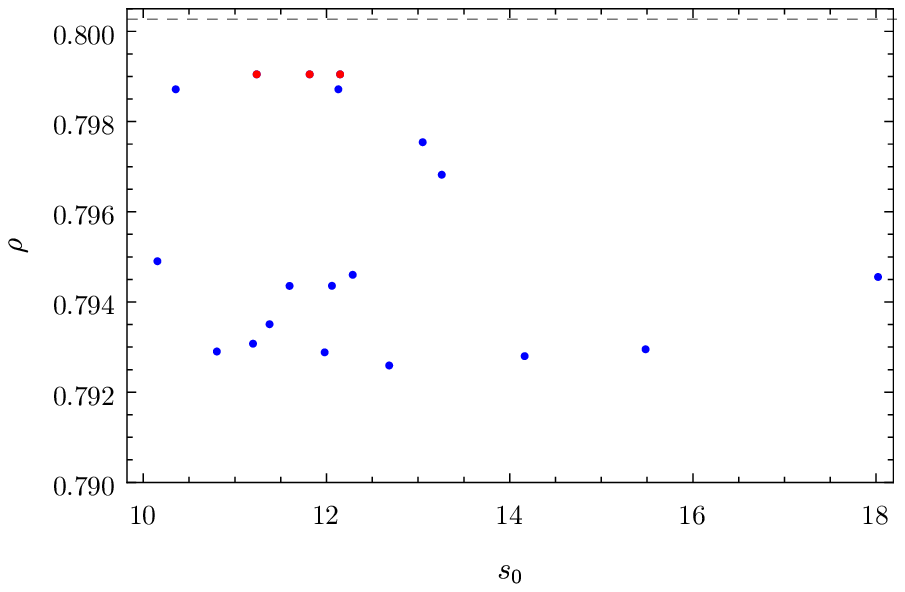}
\end{center}
\caption{Density of the final configuration as a function of a randomly chosen $s_{\rm in}$, for the case of $50$ circles using the improved algorithm (left plot) or the NÖ algorithm (right plot). The horizontal dashed line is $\rho =  	0.80027218399$, corresponding to the  currently known best packing reported in Ref.~\cite{SpechtRepo}. }
\label{Fig_1a}
\end{figure}

In Fig.~\ref{Fig_1a} we report the results of a computational experiment that we have performed
with $1000$ random trials using our improved algorithm (left plot) and the original algorithm 
of Nurmela and Östergård (right plot). In both cases the initial configurations are chosen randomly, as well as the initial value of $s$. For the plot on the left $s_{\rm in}$ has been selected randomly in the interval $(3,9)$, whereas for the plot on the right $10<s_{\rm in}<20$. In both plots only the results with density above $0.79$ are reported (blue points are used for $0.79 \leq \rho < 0.799$, red points for $0.799 \leq \rho < 0.8$ and green points for $\rho \geq 0.8$).

The different performance of the two algorithms is quite clear from the plots: in the case of our algorithm, there are $409$, $86$ and $10$ points in the three regions of density specified above; in the case of the NÖ algorithm these numbers drop to $16$, $3$ and $0$ (the highest density configuration has not been produced).

A second aspect that emerges from these plots is that $s_{\rm in} \approx 6$ is more likely to produce 
the optimal packing for the case of the present algorithm, whereas $s_{\rm in} \approx 12$ seems to produce the best (but not optimal) density for the NÖ algorithm. 

\begin{figure}
	\begin{center}
\includegraphics[width=6cm]{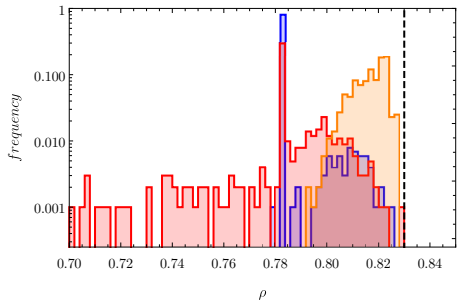}
	\end{center}
	\caption{Histograms for frequency of generating configurations with density in a given range for the case of $N=100$ circles and using $1000$ trials. The blue and red histograms are obtained using the NÖ algorithm with $s_{\rm in}=6$ (blue) and $s_{\rm in}=10$ (red); the orange histogram is the result obtained using the algorithm of the present paper with $s_{\rm in}=6$.  The dashed vertical line corresponds to the density reported in the repository of E. Specht \cite{SpechtRepo}, $\rho = 0.8300308$.}
	\label{Fig_2}
\end{figure}

In Fig.~\ref{Fig_2} we show the histograms for the frequency of generating configurations with density in a given range for the case of $N=100$ circles using $1000$ trials.
The blue and red histograms are obtained using the NÖ algorithm with $s_{\rm in}=6$ (blue) and $s_{\rm in}=10$ (red); the orange histogram is the result obtained using the algorithm of the present paper with $s_{\rm in}=6$. 
From these results, one can estimate the probability of generating a configuration with density above a certain value: for instance, using the NÖ algorithm the probability of generating a configuration with density above $0.8$ is less than $0.1$ ($0.058$ and $0.097$ for $s_{\rm in}=6$ and $s_{\rm in}=10$ respectively), whereas for the modified algorithm it is $0.993$. 
In the original scheme of Ref.~\cite{Nurmela97} the distributions are strongly peaked at a rather modest value of the density  and generating a high density configuration is extremely unlikely. 
For the modified algorithm presented in this paper, on the other hand, the distribution is moved to higher densities and spread over a larger interval. 
In this case a modest number of trials may produce nearly optimal configurations.

\begin{figure}
	\begin{center}
		\includegraphics[width=7cm]{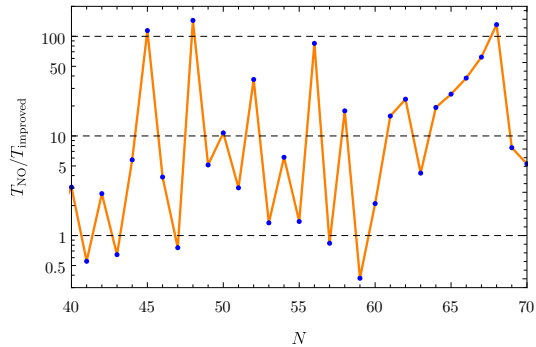}
	\end{center}
	\caption{$T_{NO}/T_{improved}$ for configurations up to $70$ circles. Higher values of the ratio mean that the improved algorithm is more effective in reaching dense configurations. }
	\label{Fig_times}
\end{figure}

Our programs run typically slightly slower for the modified algorithm than for the original algorithm, however in most cases the time required to achieve a higher density configuration is much smaller. In Fig.~\ref{Fig_times} we have compared the times taken for both algorithms (we call them $T_{\rm NO}$ and $T_{\rm improved}$ respectively) to converge to the
solutions given in Ref.~\cite{SpechtRepo} within an absolute margin of $10^{-3}$, using the same initial configurations, for $2 \leq N \leq 70$ and with a maximum number of $1000$ trials. Quite often the modified algorithm  is able to reach the solution of Ref.~\cite{SpechtRepo} even when the NÖ algorithm has not converged within $1000$ trials. In many cases there is a substantial gain in time ($T_{NO} \gg T_{\rm improved}$).

Although in our calculations with algorithm 1 we have used $s_{fin} = 10^6$, we have observed that typically for $s \approx 10^3 \leq s \leq 10^6$, the configurations change only slightly: in other words, if a configuration is not very dense at $s=10^3$ it is very unlikely that it will produce a dense configuration at $s=10^6$.  This simple observation may be used to obtain consistent gains in the time execution of the program.


\subsection{Algorithm 2}

The configurations found with the first algorithm can be improved by solving the system of nonlinear equations that correspond to the contacts between different circles and between the circles and the walls of the container. If this system has a solution, the configuration will be obtained with high precision; if the system has no solution, on the other hand, one should eliminate some of the contacts which possibly correspond to having different circles at very close distance, but not in contact. This refinement process is explained in Refs.~\cite{Peikert92,Nurmela97}.

Of course, one could  directly apply the refinement to the configurations obtained with the first algorithm, in a similar way to what done in Ref.~\cite{Nurmela97}, but  in general the packing density would only increase slightly (although the configuration would be determined very accurately). This kind of refinement is certainly preferable when the packing configuration is nearly optimal (which is more difficult to expect when $N$ is very large). 

Here we describe a second algorithm that can be applied to a densely packed configuration of circles (such as for instance those obtained using Algorithm 1) to increase the packing density (either finding a denser, different configuration or improving the current configuration).

The algorithm is composed of the following steps:
\begin{enumerate}
	\item[1)] Take a configuration of densely packed disks (for example obtained with Algorithm 1) and perturb randomly the position of the circles;
    \item[2)] Treat the centers of the circles as point--like particles repelling with an interaction $ \left(\frac{\lambda}{r^2} \right)^{s_{\rm in}}$, with $\lambda = r_{min}^2$, $r_{min}$ being the closest distance between any two points in this set;
    \item[3)] Minimize the total energy of the system, eq.~(\ref{eq_V2}), for $s=s_{\rm in}$ (typically, at the beginning we choose $s_{\rm in} \approx 10^2-10^3$);
    \item[4)] Let $s' = \kappa s$, with $\kappa > 1$, and repeat step 3) using the configuration obtained there as the initial configuration;  iterate these steps
    up to a sufficiently large value of $s$, until convergence has been reached;
    \item[5)] If the final configuration of step 4) has a higher density of the configuration of 1), use it as new initial configuration and
    repeat the steps 1--4) as many times as needed, each time updating the initial configuration to be the densest configuration; 
    \item[6)] If after some iterations the process cannot easily improve the density, modify step 1), by making the amplitude of the random perturbation smaller
    and repeat the steps 2--5) (in general the initial value of $s_{\rm in}$ will then be taken to be larger);
    \item[7)]  Stop the algorithm when convergence has been reached (or when the assigned number of iterations has been completed).
\end{enumerate}

The operation of perturbing the positions of the disks at step 1) motivates the physical analogy of improving the arrangement of a large number of small, rigid bodies inside a container by gently shaking it~\cite{Aste08}: for example,  Pouliquen, Nicolas and Weidman have shown that one can obtain dense packings of uniform spheres if the beads are poured slowly inside a container which is horizontally shaken~\cite{Pouliquen97} .  More recently, Baker and Kudrolli have performed similar experiments with beads with the shape of platonic solids~\cite{Kudrolli2010}. For this reason we will also refer to Algorithm 2 as to the ``shaking algorithm".

\section{Numerical results}
\label{sec:results}

In this section we present the numerical results obtained using the algorithms described in this paper on selected configurations~\footnote{The only criterium that we have used to select a configuration has been to require a large number of circles and that it had been obtained before.}.

In most of the cases presented here the Algorithm 1 has been applied with a limited number of trials, particularly for configurations containing a large number of circles.

In Table \ref{table_1} we report the  best densities obtained for a given $N$, with a given number of trials  and starting with  $s=s_{\rm in}$, using either the Algorithm 1 (fourth column), Algorithm 1 and 2 (fifth column) or Algorithm 2 (seventh column), using as initial configuration the corresponding configuration reported in \cite{SpechtRepo} (sixth column). Results that improve the density of \cite{SpechtRepo} are underlined. Remarkably, for all $N$ considered in the Table we have been able to improve the previous records of density.

In particular, Algorithm 2 can be used independently of Algorithm 1, as long as one can provide a sufficiently dense configuration
of disks: in general, the computational cost of applying Algorithm 2 is also smaller than for Algorithm 1. In this respect, Algorithm 2 can be used in conjunction with any packing algorithm to check whether the configuration can be improved when $N$  is very large.

\begin{table}
	\caption{Numerical results obtained with the improved algorithm (only the first 10 decimals of the results of Ref.~\cite{SpechtRepo} are reported here). The results that are underlined improve the corresponding results of \cite{SpechtRepo}.}
	\bigskip
	\label{table_1} 
	\begin{center}
		\begin{tabular}{|c|c|c|c|c|c|c|}
			\hline
			$N$ & $\#$ of & $s_{\rm in}$ & $\rho$      & $\rho$ & Ref.~\cite{SpechtRepo} & Ref.~\cite{SpechtRepo}  + \\
			    &  trials            &       & Alg. 1 & Alg. 1+2 &   & Alg. 2 \\
			\hline
			254   & 1000  & 4 & 0.849002495 & 0.849002608 & 0.8501278714 & \underline{0.8501434314}  \\
			500   & 40    & 3 & 0.861469117 & 0.863118997 & 0.8641899871 & \underline{0.8641942995} \\
			999   & 40    & 2 & \underline{0.872033110}   & \underline{0.874817581} & 0.8558502576 & - \\
			2000  & 40    & 2 & \underline{0.874958653}   & \underline{0.875264480} & 0.8639231178 & - \\
			3000  & 40    & 1 & 0.879847778 & 0.880856846 & 0.8821229717 & \underline{0.8900843290}  \\
			4000  & 40    & 1 & \underline{0.879476539}   & \underline{0.880026692} & 0.7915325405 & - \\
			9996  & 8     & 1 & \underline{0.881622922}   & \underline{0.881724377} & 0.8091772569 & - \\
			\hline
			\hline
		\end{tabular}
	\end{center}
	\bigskip\bigskip
\end{table}

\begin{figure}[H]
	\begin{center}
		\includegraphics[width=7cm]{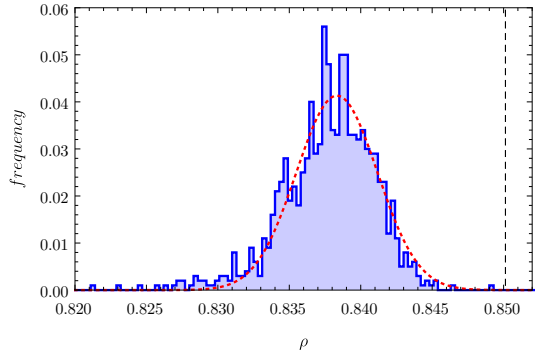} 
	\end{center}
	\caption{Histogram of the densities for $254$ circles in a square, using the improved algorithm (algorithm 1) with $s_{\rm in}=4$. The vertical dashed line in the histogram is the result reported in Ref.~\cite{SpechtRepo}, while the red dotted curve correspond to the gaussian fit ${\rm frequency}(\rho) \approx 
		\alpha \ e^{-\beta (\rho -\gamma)^2}$ with $\alpha = 4.13 \times 10^{-2}$, $\beta =6.23 \times 10^4$ and $\gamma=8.38 \times 10^{-1}$.} 
	\label{Fig_254}
\end{figure}

\begin{figure}[H]
	\begin{center}
		\includegraphics[width=5cm]{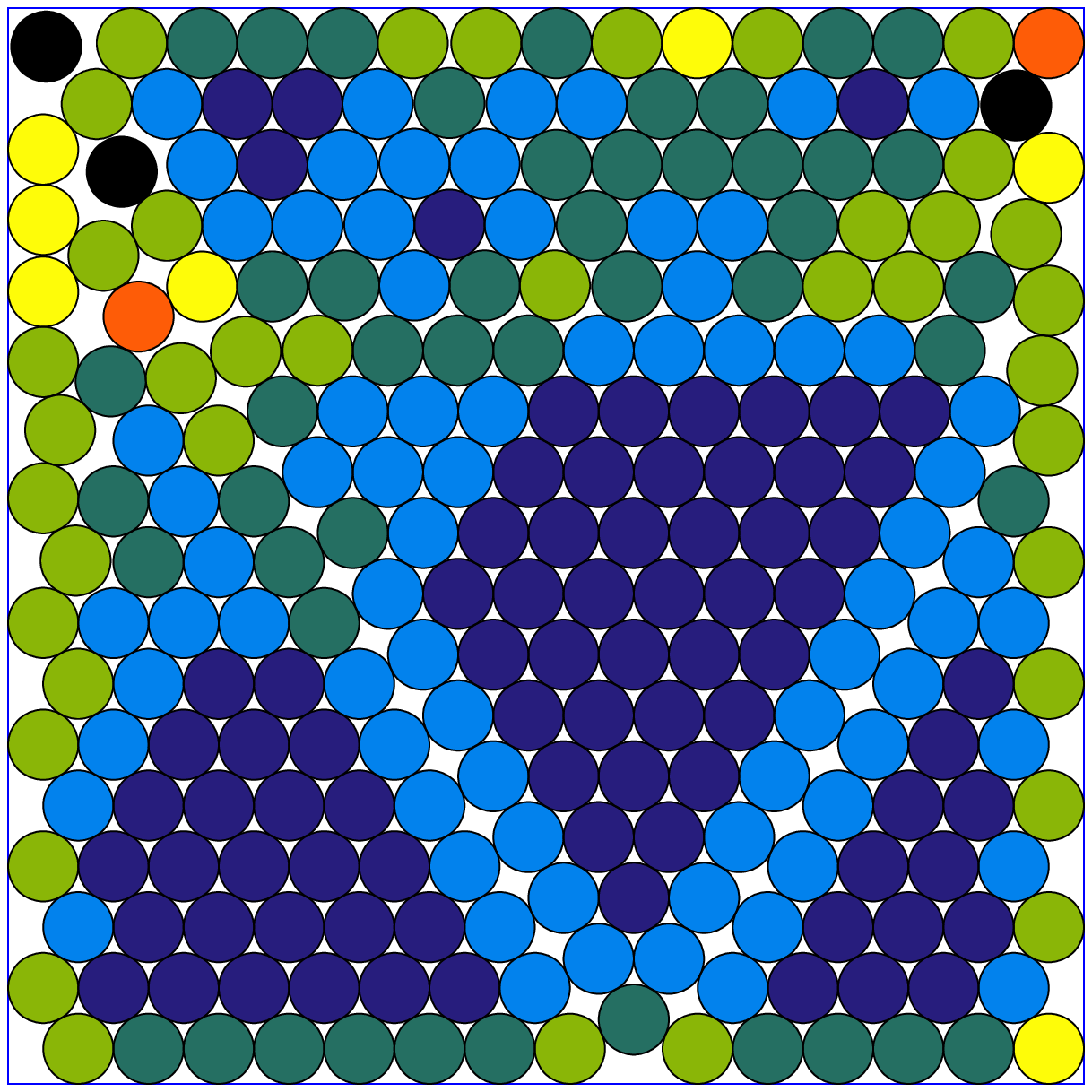}  \hspace{0.5cm}
		\includegraphics[width=5cm]{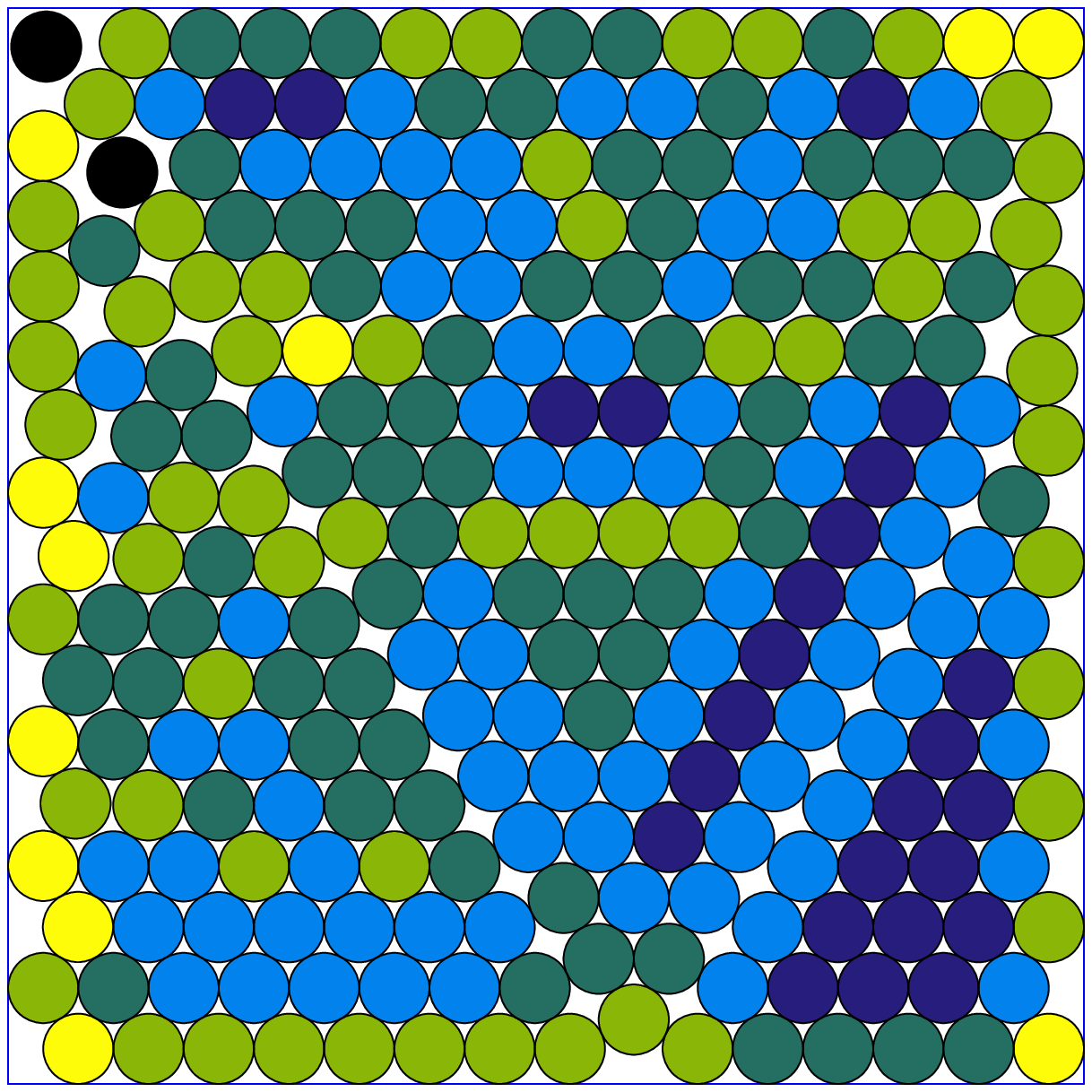}  
	\end{center}
	\caption{Left plot: Best configuration obtained for $254$ circles ($\rho = 0.8501434314$); Right plot: Best configuration reported in \cite{SpechtRepo} for $254$ circles  ($\rho = 0.8501278714$)} 
	\label{Fig_254b}
\end{figure}

In  Fig.~\ref{Fig_254}, we show the histogram of the densities obtained using Algorithm 1 with $1000$ trials. The vertical dashed line corresponds to the result reported in ref.~\cite{SpechtRepo}, while the red dotted curve correspond to the gaussian fit ${\rm frequency}(\rho) \approx \alpha \ e^{-\beta (\rho -\gamma)^2}$ with 
 $\alpha = 4.13 \times 10^{-2}$, $\beta =6.23 \times 10^4$ and $\gamma=8.38 \times 10^{-1}$~\footnote{The gaussian fit may be used to roughly estimate the number of trials needed to obtain a configuration with density above a certain value.}.

In this case, Algorithm 1 fails to improve the density reported in Ref.~\cite{SpechtRepo} with $1000$ trials and the application of Algorithm 2 to the best configuration found in this way only increases slightly the density (see columns 4 and 5 of the table); however, by applying Algorithm 2 to the configuration of \cite{SpechtRepo}, we manage to obtain a larger density (column 7 of the table). The fact the increase of the density is modest may signal that the refinement  process carried out in \cite{SpechtRepo} did not converge fully.

In the left plot of Fig.~\ref{Fig_254b} we display the configuration corresponding to this density: different colors are used to symbolize the number of closest neighbors of a given circles: black, orange, yellow, light and darker green, blue and dark blue are used to represent from $0$ to $6$ neighbors (notice however that not all colors show up in this plot). A tolerance of $10^{-4}$ is used to decide whether two circles are touching or not. A similar plot for the configuration  of \cite{SpechtRepo} is displayed in the right figure. 

For configurations that do not possess any symmetry, as in the present case, there are 8 different orientations of the points, which all correspond to the same configuration. To avoid ambiguities we have adopted the convention to choose the orientation for which the center of mass of the circles falls in the region $0 \leq \phi \leq \pi/4$~\footnote{It is easy to convince oneself that an appropriate rotation of $0$, $\pi/2$, $\pi$ or $3 \pi/2$ radians, possibly followed by a reflection about the line $y=x$, will always bring the center of mass of an arbitrary configuration to lie in this region. }. This convention is very practical at the time of comparing different configurations, as it can be appreciated from Fig.~\ref{Fig_254b}. However a purely graphical comparison may be difficult for configurations with very large number of circles and more quantitative criteria is preferable: for this reason we have defined the quantity
\begin{equation}
\upsilon = \frac{1}{N \overline{\mathcal{D}}} \sum_{j=1}^N d^{\rm (min)}_{j}
\end{equation}
where $\overline{\mathcal{D}}$ is the average diameter of the circles of the two configurations and $ d^{\rm (min)}_{j}$ is the minimal distance
of the $j^{th}$ circle of the first configuration from any of the circles of the second configuration (the two configurations need to be oriented in the canonical form). Finding $\upsilon \ll 1$ is an indication of the similarity of the two configurations: in the present case we have found $\upsilon = 0.0099911$, which supports the conclusions reached based on the graphical inspection.

The configurations in Fig.~\ref{Fig_254b}  display a large V--shaped fault along which the circles are approximately  arranged on a hexagonal lattice. Additionally, a milder horizontal fault, with two regions slightly shifted horizontally, is also observed.

The presence of domains with different arrangements, or even with the same arrangement but separated by a fault, in some way 
is similar to what is observed in the case of the Thomson problem, where the occurrence of defects is crucial in lowering the 
total energy of the system~\cite{Bowick02,Wales06,Wales14}. In absence of borders, the circles would pack more effectively by 
allowing each circle to be at the center of a  hexagonal cell, with a neighbor circle located at each of the corners of the hexagon.
It is interesting that the faults in many cases resemble the trajectory of a bouncing ball (the angles of $\pi/3$ that the ``trajectory" forms with the border allows hexagonal packing in each of the three subdomains).

In the remaining figures similar plots are displayed for configurations with $N= 500, 999, 2000, 3000, 4000$ and $9996$ circles.
For the case of $999$ we have compared three sets of results obtained with our algorithm, but with different $s_{\rm in}$, obtaining for the majority of trials better results than \cite{SpechtRepo} in all three cases.

\begin{figure}[H]
	\begin{center}
		\includegraphics[width=7cm]{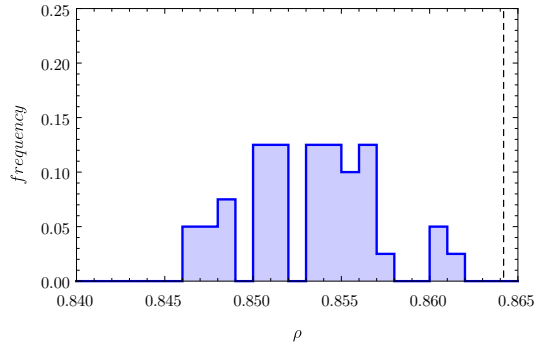} 
	\end{center}
	\caption{Histogram of the densities for $500$ circles in a square, using the Algorithm 1 with $s_{\rm in}=3$ and just $40$ trials. The vertical dashed line in the histogram is the result reported in Ref.~\cite{SpechtRepo}} 
	\label{Fig_500}
\end{figure}

\begin{figure}[H]
	\begin{center}
		\includegraphics[width=6cm]{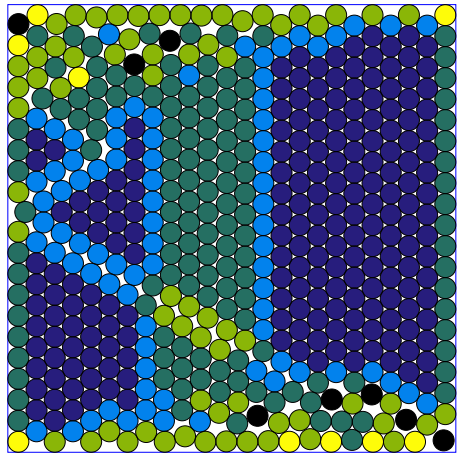}  
	\end{center}
	\caption{Best configuration obtained for $500$ circles ($\rho = 0.8641942995$). } 
	\label{Fig_500b}
\end{figure}

In Fig.~\ref{Fig_500} we display the histogram of the densities for $500$ circles in a square, using the Algorithm 1 with $s_{\rm in}=3$ and just $40$ trials. The density of the best configuration found in this way falls below the density of \cite{SpechtRepo} (although the subsequent application of Algorithm 2 is able to improve this result); as for the case of $254$ circles, when Algorithm 2 is applied to the configuration of \cite{SpechtRepo} the latter is slightly improved (see Fig.~\ref{Fig_500b}) with $\upsilon = 0.00424$ (such a small value signals that the two configurations are essentially equivalent).

\begin{figure}[H]
	\begin{center}
		\includegraphics[width=7cm]{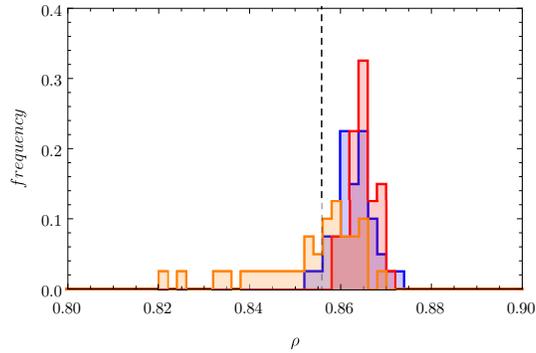} 
	\end{center}
	\caption{Histogram of densities using Algorithm 1 and  $s_{\rm in} = 2,4,6$ (blue, red and orange respectively) for  $N=999$ circles. The dashed vertical line corresponds to the density reported in the repository of E. Specht \cite{SpechtRepo}, $\rho = 0.8558502$.}
	\label{Fig_999}
\end{figure}

\begin{figure}[H]
	\begin{center}
		\includegraphics[width=6cm]{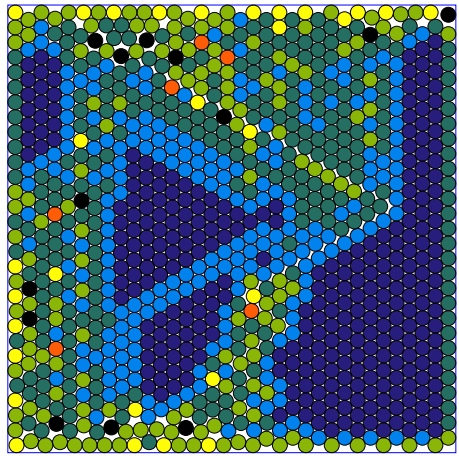}  
	\end{center}
	\caption{Best configuration obtained for $999$ circles ($\rho = 0.874817581$). } 
	\label{Fig_999b}
\end{figure}

For $999$ circles Algorithm 1 is able to improve the configuration of \cite{SpechtRepo} consistently, as one can see from 
Fig.~\ref{Fig_999}, where we plot the histogram of frequency for obtaining dense configurations using Algorithm 1 with  $s_{\rm in} = 2,4,6$ (blue, red and orange respectively) with $40$ trials for each case. By then applying Algorithm 2 to the best configuration found with Algorithm 1 we are able to still improve the density: the corresponding configuration is shown in Fig.~\ref{Fig_999b}. 
In this case the value $\upsilon = 0.70586$ confirms that our configuration is very different from that of \cite{SpechtRepo}. 

\begin{figure}[H]
	\begin{center}
		\includegraphics[width=7cm]{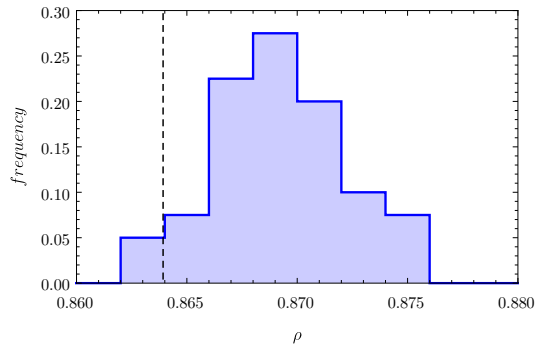} 
	\end{center}
	\caption{Histogram of the densities for $2000$ circles in a square, using Algorithm 1 with $s_{\rm in}=2$ and $40$ trials. The vertical dashed line in the histogram is the result reported in Ref.~\cite{SpechtRepo}.} 
	\label{Fig_2000}
\end{figure}

\begin{figure}[H]
	\begin{center}
		\includegraphics[width=6cm]{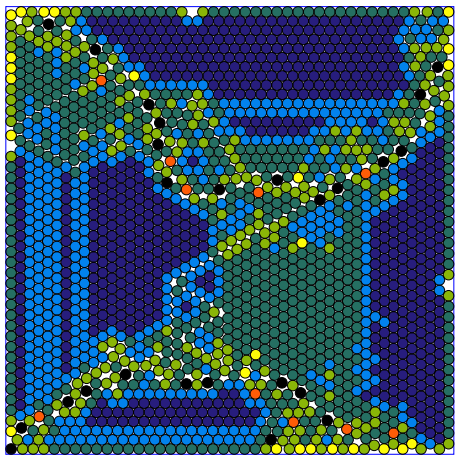}  
	\end{center}
	\caption{Best configuration obtained for $2000$ circles ($\rho = 0.875264480$). } 
	\label{Fig_2000b}
\end{figure}

\begin{figure}[H]
	\begin{center}
		\includegraphics[width=7cm]{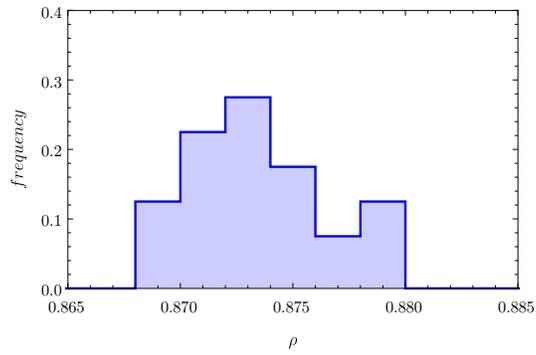} 
	\end{center}
	\caption{Histogram of the densities for $4000$ circles in a square, using Algorithm  1 with $s_{\rm in}=1$ and $40$ trials.} 
	\label{Fig_4000}
\end{figure}

\begin{figure}[H]
	\begin{center}
		\includegraphics[width=6cm]{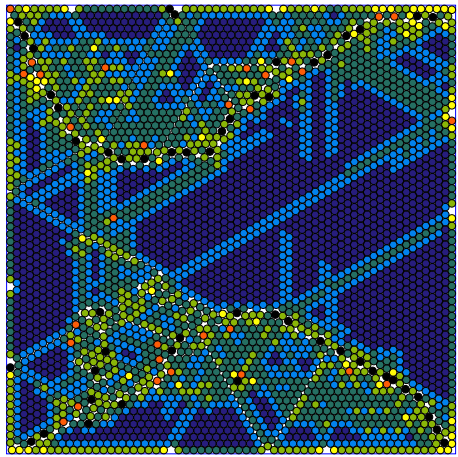}  
	\end{center}
	\caption{Best configuration obtained for $4000$ circles ($\rho = 0.880026692$). } 
	\label{Fig_4000b}
\end{figure}

\begin{figure}[H]
	\begin{center}
		\includegraphics[width=6cm]{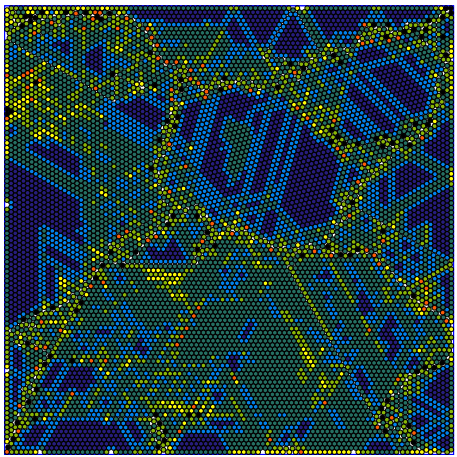}  
	\end{center}
	\caption{Best configuration obtained for $9996$ circles ($\rho = 0.881724377$). } 
	\label{Fig_9996b}
\end{figure}

The cases of  $2000$, $4000$ and $9996$ circles are similar to the one of $999$, in the sense that Algorithm 1 is able to find a better configuration than the one of Ref.~\cite{SpechtRepo}, and the subsequent application of Algorithm 2 further improves the density (see Figs.~\ref{Fig_2000}, \ref{Fig_2000b}, \ref{Fig_4000}, \ref{Fig_4000b} and \ref{Fig_9996b}).

\begin{figure}[H]
	\begin{center}
		\includegraphics[width=5cm]{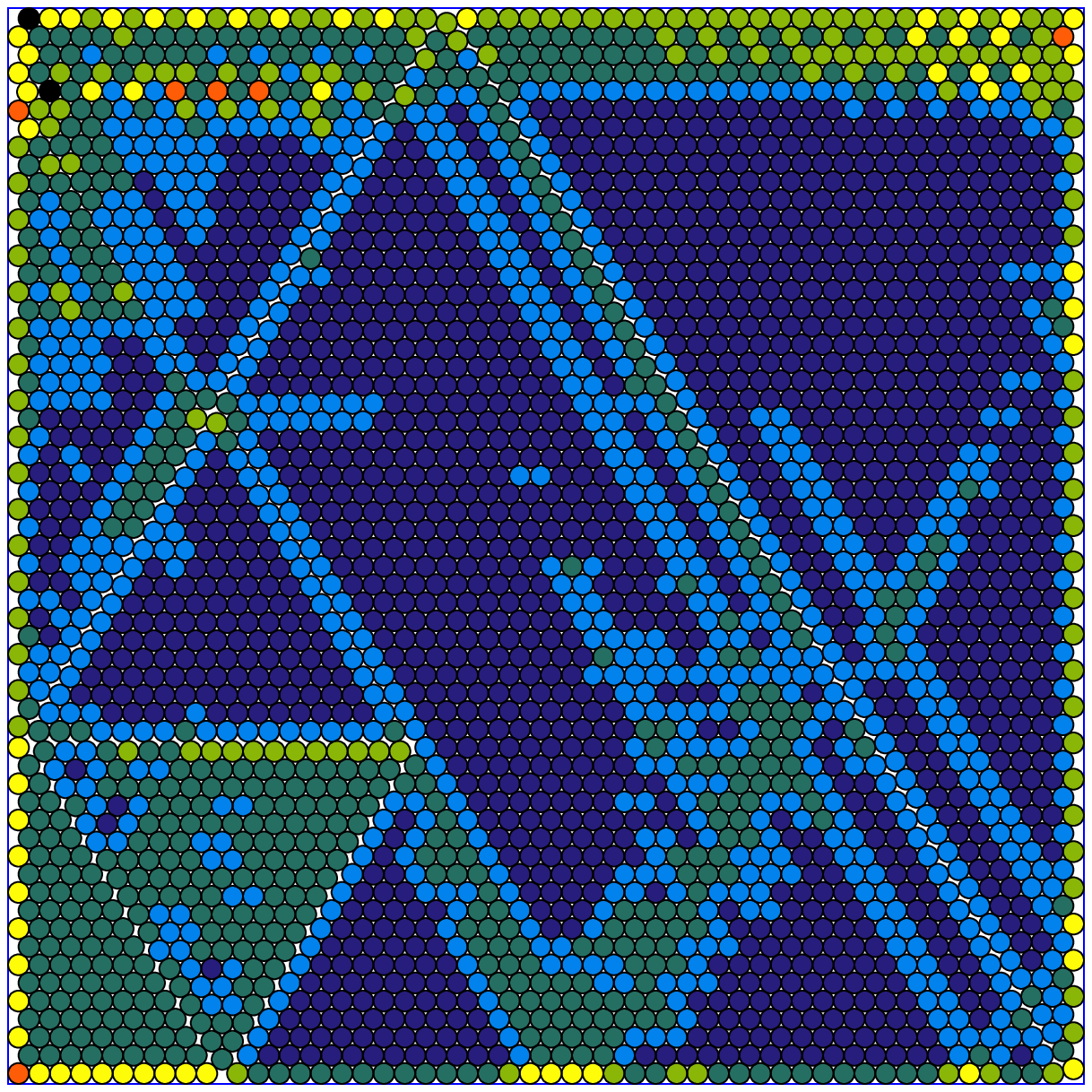}   \hspace{0.5cm}
		\includegraphics[width=5cm]{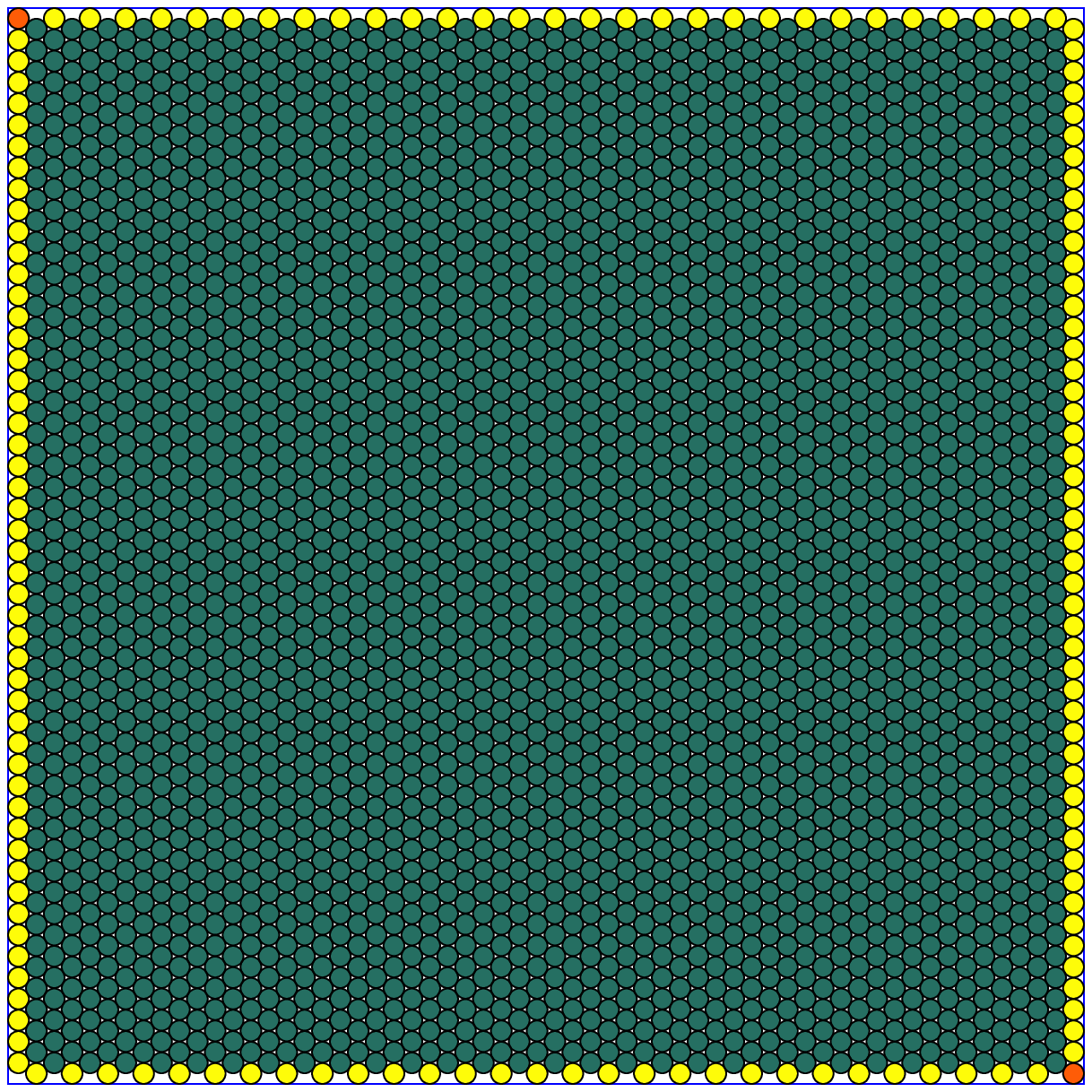}  
	\end{center}
	\caption{Left plot: Best configuration obtained for $3000$ circles ($\rho = 0.8900843290$); Right plot: Configuration reported in Ref.~\cite{SpechtRepo}, ($\rho = 0.8821229717$).} 
	\label{Fig_3000b}
\end{figure}

The case of $3000$ circles is more interesting since the configuration of Ref.~\cite{SpechtRepo} is both dense and regular, as shown in the right plot in Fig.~\ref{Fig_3000b} (it might even appear to be a good candidate for a global maximum of the density). Algorithm 1 is not able to improve the result of Ref.~\cite{SpechtRepo} in $40$ trials, although getting rather close; Algorithm 2, does improve the configuration obtained with Algorithm 1, but again it does not overcome  Ref.~\cite{SpechtRepo}. By applying Algorithm 2 directly to the configuration of  Ref.~\cite{SpechtRepo}, however,  we obtain a denser (and less regular) configuration: it is interesting to observe that in this case faults resembling 
a bouncing ball trajectory appear (Fig.~\ref{Fig_3000b}, left).

\begin{figure}[H]
	\begin{center}
		\includegraphics[width=7cm]{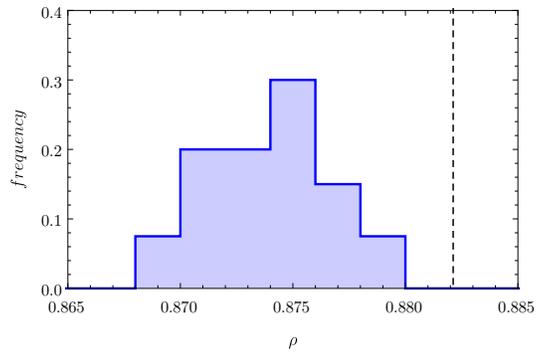} 
	\end{center}
	\caption{Histogram of the densities for $3000$ circles in a square, using Algorithm 1 with $s_{\rm in}=1$ and $40$ trials. The vertical dashed line in the histogram is the result reported in Ref.~\cite{SpechtRepo}} 
	\label{Fig_3000}
\end{figure}

As a final test of the effectiveness of Algorithm 2 we have applied it to the configurations reported in \cite{SpechtRepo} for $N \leq 650$, using $10$ trials for each $N$. The densities of the configurations that have been improved are reported in the Tables \ref{table_2}, \ref{table_3} and \ref{table_4}.

\begin{table}
\caption{Configurations from Ref.~\cite{SpechtRepo} for $N \leq 500$ improved with an application of Algorithm 2 with $10$ trials.}
\bigskip
\label{table_2} 
\begin{center}
\begin{tabular}{|c|c|c|c|}
\hline
$N$ & Ref.~\cite{SpechtRepo} & Ref.~\cite{SpechtRepo} +  Alg. 2  & $\delta\rho$  \\
\hline
171   &  0.837580192489   & 0.837580671700 & $4.7921 \times 10^{-7}$ \\
220   &  0.846035170012   & 0.846035308218 & $1.3821 \times 10^{-7}$ \\
224   &  0.848926223422   & 0.848926484157 & $2.6074 \times 10^{-7}$ \\     
244   &  0.856614482106   & 0.856621929790 & $7.4477 \times 10^{-6}$ \\
249   &  0.850028432564   & 0.850029524580 & $1.0920 \times 10^{-6}$ \\
251   &  0.846860216473   & 0.846868469652 & $8.2531 \times 10^{-6}$ \\
252   &  0.847905012837   & 0.847942109618 & $0.00003709678$ \\
254   &  0.850127871456   & 0.850140054769 & $0.00001218331$ \\
259   &  0.848279073344   & 0.848282532997 & $3.45965 \times 10^{-6}$ \\
260   &  0.849846539415   & 0.849847399152 & $8.5974 \times 10^{-7}$ \\
288   &  0.852791129258   & 0.852810840856 & $0.00001971160$ \\
308   &  0.857474722364   & 0.857475077601 & $3.5524 \times 10^{-7}$ \\
320   &  0.854435705036   & 0.854437153145 & $1.4481 \times 10^{-6}$ \\
326   &  0.858529615776   & 0.858531735939 & $2.1202 \times 10^{-6}$ \\
342   &  0.859651812122   & 0.859655592894 & $3.7808 \times 10^{-6}$ \\
343   &  0.857711713800   & 0.857730279172 & $0.00001856537$ \\
350   &  0.859297933840   & 0.859299191735 & $1.2578 \times 10^{-6}$ \\
354   &  0.857808914126   & 0.857817899935 & $8.9858 \times 10^{-6}$ \\
355   &  0.855876794760   & 0.855876794919 & $1.6 \times 10^{-10}$ \\
360   &  0.857669063503   & 0.857669322453 & $2.5895 \times 10^{-7}$ \\
366   &  0.865471406680   & 0.865472365965 & $9.5928 \times 10^{-7}$ \\
374   &  0.864779112892   & 0.864780351180 & $1.2383 \times 10^{-6}$ \\
379   &  0.861841567643   & 0.861850960552 & $9.3929 \times 10^{-6}$ \\
381   &  0.858992669627   & 0.858992826227 & $1.5660 \times 10^{-7}$ \\ 
386   &  0.859803437278   & 0.859814435704 & $0.00001099843$ \\
387   &  0.860124928457   & 0.860125172298 & $2.4384 \times 10^{-7}$ \\
388   &  0.860600088374   & 0.860603124621 & $3.0362 \times 10^{-6}$ \\ 
390   &  0.862147420450   & 0.862152329448 & $4.9090 \times 10^{-6}$ \\
394   &  0.860340577446   & 0.860341008748 & $4.3130 \times 10^{-7}$ \\
398   &  0.862130776263   & 0.862132631510 & $1.8552 \times 10^{-6}$ \\
399   &  0.861313883465   & 0.861314923791 & $1.0403 \times 10^{-6}$ \\
404   &  0.867916086699   & 0.867916814994 & $7.2829 \times 10^{-7}$ \\
416   &  0.864618368821   & 0.864618745652 & $3.7683 \times 10^{-7}$ \\
420   &  0.860437672287   & 0.860446345152 & $8.6729 \times 10^{-6}$ \\
429   &  0.861584962241   & 0.861590828839 & $5.8666 \times 10^{-6}$ \\
430   &  0.862535308662   & 0.862535428943 & $1.2028 \times 10^{-7}$ \\
432   &  0.862874861575   & 0.862875512999 & $6.5142 \times 10^{-7}$ \\
441   &  0.867073370478   & 0.867074491295 & $1.1208 \times 10^{-6}$ \\
442   &  0.868136259241   & 0.868142323692 & $6.0644 \times 10^{-6}$ \\
446   &  0.872846723718   & 0.872846820520 & $9.680 \times 10^{-8}$ \\
454   &  0.863683581044   & 0.863688552873 & $4.9718 \times 10^{-6}$ \\
455   &  0.863342850046   & 0.863374076458 & $0.00003122641$ \\
457   &  0.863542171511   & 0.863543941143 & $1.7696 \times 10^{-6}$ \\
459   &  0.866500150886   & 0.866500438923 & $2.8804 \times 10^{-7}$ \\
460   &  0.866490384537   & 0.866493363775 & $2.9792 \times 10^{-6}$ \\
463   &  0.862438056016   & 0.862447838092 & $9.7821 \times 10^{-6}$ \\
464   &  0.860522797002   & 0.860522910370 & $1.1337 \times 10^{-7}$ \\
466   &  0.861274479076   & 0.861287539763 & $0.00001306069$ \\
468   &  0.863667323538   & 0.863670395255 & $3.0717 \times 10^{-6}$ \\
471   &  0.863527572747   & 0.863531796352 & $4.2236 \times 10^{-6}$ \\
479   &  0.869368037362   & 0.869392338620 & $0.00002430126$ \\
483   &  0.869954923724   & 0.869955489705 & $5.6598 \times 10^{-7}$ \\
494   &  0.866506985501   & 0.866507633073 & $6.4757 \times 10^{-7}$ \\
495   &  0.865134182063   & 0.865161262403 & $0.00002708034$ \\
496   &  0.864513626527   & 0.864531995490 & $0.00001836896$ \\
500   &  0.864189987393   & 0.864194836255 & $4.84886 \times 10^{-6}$ \\
\hline
\hline
\end{tabular}
\end{center}
\bigskip\bigskip
\end{table}

\begin{table}
\caption{Configurations from Ref.~\cite{SpechtRepo} for $501 \leq N \leq 600$ improved with an application of Algorithm 2 with $10$ trials.}
\bigskip
\label{table_3} 
\begin{center}
\begin{tabular}{|c|c|c|c|}
\hline
$N$ & Ref.~\cite{SpechtRepo} & Ref.~\cite{SpechtRepo} +  Alg. 2  & $\delta\rho$  \\
\hline
501   &  0.863889382213   & 0.863961877953 & $0.00007249574$ \\
503   &  0.866217425179   & 0.866217758902 & $3.3372 \times 10^{-7}$ \\
505   &  0.867316099804   & 0.867322232190 & $6.1324 \times 10^{-6}$ \\
507   &  0.862650109591   & 0.862714364404 & $0.00006425481$ \\
510   &  0.863562744209   & 0.863599965836 & $0.00003722163$ \\
512   &  0.865952634824   & 0.865959090903 & $6.4561 \times 10^{-6}$ \\
515   &  0.866484772738   & 0.866704644738 & $0.00021987200$ \\
516   &  0.865335507777   & 0.865353004330 & $0.00001749655$ \\
520   &  0.868978862399   & 0.868980903951 & $2.0416 \times 10^{-6}$ \\
522   &  0.870666073669   & 0.870666905880 & $8.3221 \times 10^{-7}$ \\
523   &  0.872287882407   & 0.872288247418 & $3.6501 \times 10^{-7}$ \\
526   &  0.871878031808   & 0.871886240061 & $8.2082 \times10^{-6}$ \\
529   &  0.873250099801   & 0.873288656204 & $0.00003855640$ \\
531   &  0.871839013521   & 0.871854186814 & $0.00001517329$ \\
533   &  0.869967846300   & 0.869988326597 & $0.00002048030$ \\
535   &  0.871421752958   & 0.871457251908 & $0.00003549895$ \\
536   &  0.872241196392   & 0.872242423102 & $1.2267 \times 10^{-6}$ \\
537   &  0.873274783462   & 0.873301457473 & $0.00002667401$ \\
540   &  0.865625473043   & 0.865625825430 & $3.5239 \times 10^{-7}$ \\
541   &  0.865230368435   & 0.865251512143 & $0.00002114371$ \\
542   &  0.865029762791   & 0.865036520883 & $6.7581 \times 10^{-6}$ \\
543   &  0.863373299076   & 0.863379080351 & $5.7813 \times 10^{-6}$ \\
544   &  0.864285882153   & 0.864293252077 & $7.3699 \times 10^{-6}$ \\
545   &  0.865265664519   & 0.865266317801 & $6.5328 \times 10^{-7}$ \\
546   &  0.865463712225   & 0.865523571898 & $0.00005985967$ \\
549   &  0.867524664577   & 0.867526758525 & $2.0940 \times 10^{-6}$ \\
551   &  0.867627831331   & 0.867683111293 & $0.00005527996$ \\
553   &  0.866578862436   & 0.866579078126 & $2.1569 \times 10^{-7}$ \\
554   &  0.865820553042   & 0.865870132338 & $0.00004957930$ \\
555   &  0.866053327058   & 0.866053566749 & $2.3969 \times 10^{-7}$ \\
556   &  0.865988239406   & 0.866029246269 & $0.00004100686$ \\
557   &  0.867328052036   & 0.867328099298 & $1.2837 \times 10^{-6}$ \\
562   &  0.869169446481   & 0.869170730181 & $1.2445 \times 10^{-6}$ \\
563   &  0.869333992900   & 0.869342222812 & $8.2299 \times 10^{-6}$ \\
564   &  0.869999841216   & 0.870002205900 & $2.3647 \times 10^{-6}$ \\
565   &  0.870222707754   & 0.870223407984 & $7.0023 \times 10^{-7}$ \\
566   &  0.871338504329   & 0.871341906898 & $3.4026 \times 10^{-6}$ \\
567   &  0.872579777697   & 0.872580909689 & $1.1320 \times 10^{-6}$ \\ 
568   &  0.873201953504   & 0.873261436878 & $0.0000594834$ \\
569   &  0.874450946637   & 0.874466541377 & $0.00001559474$ \\
570   &  0.875902723385   & 0.875902934501 & $2.1112 \times 10^{-7}$  \\
573   &  0.870243195381   & 0.870266324446 & $0.00002312906$ \\
574   &  0.870014810673   & 0.870158069254 & $0.0001432585$ \\
575   &  0.870713577974   & 0.870914661306 & $0.00020108333$ \\
576   &  0.871561678533   & 0.871596945475 & $0.00003526694$ \\
577   &  0.870807261262   & 0.870815032352 & $7.7711 \times 10^{-6}$ \\
578   &  0.870327332550   & 0.870357293623 & $0.00002996107$ \\
580   &  0.868819041117   & 0.868822821814 & $3.7807 \times 10^{-6}$ \\
581   &  0.869690432018   & 0.869690677284 & $2.4527 \times 10^{-7}$ \\
582   &  0.869834042797   & 0.869861032346 & $0.00002698955$ \\
583   &  0.871237224770   & 0.871238068473 & $8.4370 \times 10^{-7}$ \\
584   &  0.871269454741   & 0.871487535310 & $0.00021808057$ \\
588   &  0.866259826278   & 0.866259945606 & $1.1933 \times 10^{-7}$ \\
589   &  0.865828323308   & 0.866030784874 & $0.00020246157$ \\
590   &  0.865455529211   & 0.865456304857 & $7.7565 \times 10^{-7}$ \\
591   &  0.865689054855   & 0.865694114205 & $5.0594 \times 10^{-6}$ \\
592   &  0.865915651949   & 0.865918963816 & $3.3119 \times 10^{-6}$ \\
593   &  0.866474107729   & 0.866520562870 & $0.0000464551$ \\
596   &  0.869538878790   & 0.869544545363 & $5.6666 \times 10^{-6}$ \\
597   &  0.867672537804   & 0.867680366507 & $7.8287 \times 10^{-6}$ \\
599   &  0.869162555009   & 0.869187481810 & $0.0000249268$ \\
600   &  0.869649946316   & 0.869654568950 & $4.6226 \times 10^{-6}$ \\
\hline
\hline
\end{tabular}
\end{center}
\bigskip\bigskip
\end{table}

\begin{table}
\caption{Configurations from Ref.~\cite{SpechtRepo} for $601 \leq N \leq 700$ improved with an application of Algorithm 2 with $10$ trials.}
\bigskip
\label{table_4} 
\begin{center}
\begin{tabular}{|c|c|c|c|}
\hline
$N$ & Ref.~\cite{SpechtRepo} & Ref.~\cite{SpechtRepo} +  Alg. 2  & $\delta\rho$  \\
\hline
601   &  0.869420245190   & 0.869423265696 & $3.0205 \times 10^{-6}$ \\
602   &  0.869868732386   & 0.869868738401 & $6.01 \times 10^{-9}$ \\
603   &  0.869837128266   & 0.869838026368 & $8.9810 \times 10^{-7}$ \\
604   &  0.869951286405   & 0.869953360859 & $2.0744 \times 10^{-6}$ \\
605   &  0.871079491309   & 0.871097664743 & $0.00001817343$ \\
606   &  0.872097795344   & 0.872136767186 & $0.00003897184$ \\
607   &  0.872527879524   & 0.872528010521 & $1.3100 \times 10^{-7}$ \\
611   &  0.872499479424   & 0.872503003875 & $3.52445 \times 10^{-6}$ \\
612   &  0.873459221940   & 0.873480604536 & $0.00002138260$ \\
615   &  0.874709672377   & 0.874736354158 & $0.00002668178$ \\
616   &  0.875773826171   & 0.875776991002 & $3.1648 \times 10^{-6}$ \\
617   &  0.876863623193   & 0.876864098581 & $4.7539 \times 10^{-7}$ \\
618   &  0.878117772912   & 0.878119225152 & $1.4522 \times 10^{-6}$ \\
619   &  0.879452627629   & 0.879452928629 & $3.0100 \times 10^{-7}$ \\
620   &  0.880847904124   & 0.880849993453 & $2.0893 \times 10^{-6}$ \\
623   &  0.870203752152   & 0.870229319512 & $0.00002556736$ \\
624   &  0.869872506030   & 0.869875855771 & $3.34974 \times 10^{-6}$ \\
625   &  0.870125767955   & 0.870159930736 & $0.00003416278$ \\
626   &  0.869802295586   & 0.870728607141 & $0.00092631155$ \\
627   &  0.870127086335   & 0.870382718996 & $0.00025563266$ \\
628   &  0.869569416951   & 0.869613705613 & $0.00004428866$ \\
629   &  0.867935069145   & 0.868144574804 & $0.000209506$ \\
630   &  0.870295590263   & 0.870297070507 & $1.4802 \times 10^{-6}$ \\
631   &  0.870830292494   & 0.870832305456 & $2.0130 \times 10^{-6}$ \\
632   &  0.870229342639   & 0.870256642728 & $0.0000273001$ \\
634   &  0.871147958848   & 0.871151359771 & $3.4009 \times 10^{-6}$ \\
635   &  0.872449352552   & 0.872469872978 & $0.00002052043$ \\
636   &  0.871027345995   & 0.871041389690 & $0.00001404369$ \\
637   &  0.868307107702   & 0.868323047165 & $0.00001593946$ \\
638   &  0.868285154330   & 0.868347478252 & $0.00006232392$ \\
639   &  0.868504832648   & 0.868505202255 & $3.6961 \times 10^{-7}$ \\ 
640   &  0.868423583375   & 0.868423860502 & $2.7713 \times 10^{-7}$ \\
641   &  0.867554575537   & 0.867569095503 & $0.00001451997$ \\
642   &  0.868184433359   & 0.868185085751 & $6.5234 \times 10^{-7}$ \\
643   &  0.869060723899   & 0.869064845986 & $4.1221 \times 10^{-6}$  \\
644   &  0.870128248670   & 0.870128264088 & $1.542 \times 10^{-8}$ \\
649   &  0.869668414889   & 0.869688086675 & $0.00001967179$ \\
\hline
\hline
\end{tabular}
\end{center}
\bigskip\bigskip
\end{table}

As a technical note, most of the calculations have been carried out on a computer AMD Ryzen 9 3950x with 16 double core processors and on an Intel Xeon E5-2640 with $8$ double core processors. On the first machine 
the average time (in seconds) needed to obtain one configuration with Algorithm 1  is approximately described by the fit 
$-0.234 N + 0.0101 N^2$.

Both algorithms have been implemented in python, using the numba compiler~\cite{numba}, which allows to consistenly speed up the calculations with respect to pure python and make it competitive with C++ (see ref.~\cite{Fua20} for a comparison of C++ and numba for the N-queens problem).

The data of the configurations obtained in this paper can be found in the Colima Repository of Computational Physics and Applied Mathematics (CRCPAM) ~\cite{crcpam}.

\section{Conclusions}
\label{sec:conclusions}
Finding a nearly optimal packing of $N$ equal circles inside a square is a challenging task: as a matter of fact, proofs of optimality exist only for configurations of up to $33$ and for $36$ circles,  while obtaining  numerically good candidates to optimal solutions is also very demanding, given the proliferation of possible configurations with very similar density, for $N \gg 1$ (Ref.~\cite{SpechtRepo} reports the current best numerical  results for circle packing in a square).

In this paper we have introduced two algorithms that can produce dense packing configurations of congruent circles inside a square. The effectiveness of the algorithms is shown by running numerical experiments for different $N$ and improving the best results in the literature. Finding the configurations corresponding to a global maximum of the packing density can be a prohibitive task for the systems of the size that we are considering here (up to several thousands disks), but at least we provide an efficient computational tool to produce highly dense packings.

The algorithms that we have introduced  can be described as an example of {\sl basin-hopping with threshold acceptance}, as the 
configurations that do not improve the density are rejected. An alternative direction that could be interesting to explore 
in the future would be introducing a fictitious temperature in the problem, that would allow to sample the low--lying solutions
at least for low temperatures~\cite{Wales21}. 

The extension of the methods discussed in the present paper to more general packing problems is currently underway. 
It is worth saying that the fundamental idea of our Algorithm 1 (i.e. border repulsion) can be applied to systems with a different interaction than the contact interaction corresponding to a packing problem. For the case of Coulomb repulsion, for instance, one is dealing with a Thomson problem in a square (or any other domain for that matter) and the computational gains introduced by the border repulsion should still be present in this case. In ref.~\cite{Amore16}, which deals with the Thomson problem inside a circle and inspired the original idea for the present work, one of us observed that the path to reaching optimal solution is definitely shortened by controlling (although in a different way) the number of peripheral charges. Actually, we expect that the effectiveness of our method for packing should drastically increase for packing problems in three dimensions: because of Gauss' law, the charges distribute on the surface of a conductor, and therefore the NÖ algorithm should possibly need to use larger values of $s_{\rm in}$, compared to its two--dimensional analog.
However, using larger values of $s_{\rm in}$ is likely to affect the ability to select a tight arrangement of spheres inside the volume. No such problem is expected in our algorithm, where the screening of the border in the early stages allows it to work with modest values of $s_{\rm in}$ and thus effectively reach dense configurations.

\section*{Acknowledgements}
The authors would like to thank Dr Roberto S\'aenz, Dr Eckard Specht and Dr David Wales, for reading the manuscript and  for their valuable comments.  The research of P.A. was supported by the Sistema Nacional de Investigadores (M\'exico). P.A. would like to thank the Universidad de Colima, and in particular M.C. Miguel Angel Rodriguez Ortiz for the support in the creation of the Repository of Computational Physics and Applied Mathematics at the Universidad de Colima. 
The plots in this paper have been plotted using {\rm MaTeX} \cite{szhorvat} and Mathematica \cite{wolfram}.

\section*{Conflict of interest}

 The authors declare that they have no conflict of interest.


\begin{thebibliography}{}
\bibitem{FejesToth42} Fejes, L. ``Über die dichteste Kugellagerung", Mathematische Zeitschrift 48.1 (1942): 676-684.
\bibitem{Schaer65} Schaer, J. ``The densest packing of 9 circles in a square", Canadian Mathematical Bulletin 8.3 (1965): 273-277.
\bibitem{Schaer65b} Schaer, J., and A. Meir. ``On a geometric extremum problem'', Canadian Mathematical Bulletin 8.1 (1965): 21-27.
\bibitem{Szabo07b} Szab\'o, P. G., et al. ``New approaches to circle packing in a square: with program codes", Vol. 6. Springer Science \& Business Media, 2007.
\bibitem{Markot21} Mark\'ot, M. C., ``Improved interval methods for solving circle packing problems in the unit square." Journal of Global Optimization 81.3 (2021): 773-803.
\bibitem{Hifi09} Hifi, M., and Rym M., ``A literature review on circle and sphere packing problems: Models and methodologies", Advances in Operations Research 2009 (2009).
\bibitem{SpechtRepo} Specht, E., \\ \href{http://hydra.nat.uni-magdeburg.de/packing/}{Packings of equal and unequal circles in fixed-sized containers $\dots$}, {\sl accessed on Oct. 2, 2020}
\bibitem{Nurmela97} Nurmela, K. J., and Östergård, P. , ``Packing up to 50 equal circles in a square", Discrete \& Computational Geometry 18.1 (1997): 111-120.
\bibitem{Hales05} Hales, T. C., ``A proof of the Kepler conjecture", Annals of mathematics (2005): 1065-1185.
\bibitem{Stokely03} Stokely, K., Ari D., and Scott V. F., ``Two-dimensional packing in prolate granular materials", Physical Review E 67.5 (2003): 051302.
\bibitem{Salerno18} Salerno, K. M., et al., ``Effect of shape and friction on the packing and flow of granular materials", Physical Review E 98.5 (2018): 050901.
\bibitem{deBono20} de Bono, J. P., and McDowell, G.R., ``On the packing and crushing of granular materials", International Journal of Solids and Structures 187 (2020): 133-140.
\bibitem{Erber91} Erber, T., and Hockney, G. M. , ``Equilibrium configurations of N equal charges on a sphere", Journal of Physics A: Mathematical and General 24.23 (1991): L1369.
\bibitem{Perez96} Pérez-Garrido, A., et al., ``Many-particle jumps algorithm and Thomson's problem", Journal of Physics A: Mathematical and General 29.9 (1996): 1973. 
\bibitem{Bowick02} Bowick, M. et al., ``Crystalline order on a sphere and the generalized Thomson problem", Physical Review Letters 89.18 (2002): 185502. 
\bibitem{Wales06} Wales, D. J., and Ulker, S. , ``Structure and dynamics of spherical crystals characterized for the Thomson problem", Physical Review B 74.21 (2006): 212101.
\bibitem{Wales14} Wales, D. J  ``Chemistry, geometry, and defects in two dimensions", ACS nano 8.2 (2014): 1081-1085.
\bibitem{Amore16} Amore, P., ``Comment on 'Thomson rings in a disk' ", Physical Review E 95.2 (2017): 026601.	
\bibitem{Peikert92} Peikert, R. et al., ``Packing circles in a square: a review and new results", System modelling and optimization. Springer, Berlin, Heidelberg, 1992. 45-54.
\bibitem{Aste08} Weaire, D., and Aste T.,  ``The pursuit of perfect packing". CRC Press, 2008.
\bibitem{Pouliquen97} Pouliquen, O. , Maxime N., and Weidman, P. D. , ``Crystallization of non-Brownian spheres under horizontal shaking", Physical Review Letters 79.19 (1997): 3640.
\bibitem{Kudrolli2010} Baker J. and Kudrolli, A., ``Maximum and minimum stable random packings of platonic solids", Physical Review E 82.6 (2010): 061304.		
\bibitem{numba} Lam, S. K., Antoine P., and Seibert S., ``Numba: A llvm-based python jit compiler." Proceedings of the Second Workshop on the LLVM Compiler Infrastructure in HPC. (2015).
\bibitem{Fua20} Fua, P., and Lis K., "Comparing python, go, and c++ on the n-queens problem." arXiv preprint arXiv:2001.02491 (2020).
\bibitem{crcpam} Amore, P., \\
{\small \href{https://ucrcpam.ucol.mx/}{``Universidad de Colima Repository of Computational Physics and Applied Mathematics"}} (2021)
\bibitem{Wales21} Wales, D. J-, Private communication (2021)
\bibitem{szhorvat} Szabolcs H., "LaTeX typesetting in Mathematica",  http://szhorvat.net/pelican/latex-typesetting-in-mathematica.html
\bibitem{wolfram} Wolfram Research, Inc., Mathematica, Version 12.3.1, Champaign, IL (2021).
\end{thebibliography}
\end{document}